\documentclass[12pt,showpacs,preprintnumbers,amsmath,amssymb]{revtex4}
\usepackage{graphicx}
\usepackage{epsfig}
\usepackage{dcolumn}
\usepackage{bm}
\usepackage{url}
\begin{document}

\title{\bf Update of the $e^+e^-\to\pi^+\pi^-$ cross section measured by SND
           detector in the energy region $400<\sqrt[]{s}<1000$ MeV.}
\author{ M.N.Achasov}\email{achasov@inp.nsk.su}
\author{ K.I.Beloborodov} 
\author{A.V.Berdyugin}
\author{A.G.Bogdanchikov}
\author{A.V.Bozhenok} 
\author{A.D.Bukin}
\author{D.A.Bukin}
\author{T.V.Dimova} 
\author{V.P.Druzhinin} 
\author{V.B.Golubev} 
\author{A.A.Korol} 
\author{S.V.Koshuba}
\author{E.V.Pakhtusova} 
\author{S.I.Serednyakov} 
\author{Yu.M.Shatunov}
\author{V.A.Sidorov}
\author{Z.K.Silagadze} 
\author{A.N.Skrinsky}
\author{Yu.A.Tikhonov}
\author{A.V.Vasiljev}
\affiliation{ 
         Budker Institute of Nuclear Physics,  \\
         Siberian Branch of the Russian Academy of Sciences \\
         11 Lavrentyev,Novosibirsk,630090, Russia \\
	 Novosibirsk State University, \\
         630090, Novosibirsk, Russia}

\begin{abstract}
 The corrected cross section of the  $e^+e^-\to \pi^+\pi^-$ process measured in
 the SND experiment at the VEPP-2M $e^+e^-$ collider is presented. The update
 is necessary due to a flaw in the $e^+e^-\to\pi^+\pi^-$ and 
 $e^+e^-\to\mu^+\mu^-$ Monte Carlo events generators used previously in data 
 analysis.
\end{abstract}


\maketitle

 The spherical neutral detector SND \cite{sndnim} operated from 1995 to 2000 at
 VEPP-2M $e^+e^-$ collider \cite{vepp2}. One of the recent SND results was the
 measurement of the $e^+e^-\to\pi^+\pi^-$ process cross
 section in the energy region $\sqrt{s}<1000$ MeV \cite{sndpipi}. The
 systematic error of the cross section determination was estimated to be 
 1.3 \%. Studies of the $e^+e^-\to\pi^+\pi^-$ reaction allow us to determine 
 the $\rho$ and $\omega$ meson parameters and provide information on the 
 $G$-parity violation mechanism in the $\omega\to\pi^+\pi^-$ decay.

 In the last time the comprehension of the high precision results of the muon 
 anomalous magnetic moment measurement \cite{bnl1,bnl2} attracted heightened 
 attention to the $e^+e^-\to\pi^+\pi^-$ cross section. 
 
 The comparison of the $e^+e^-\to\pi^+\pi^-$ process cross section with the 
 spectral function in  the $\tau^\pm\to\pi^\pm\pi^0\nu_\tau$ decay 
 \cite{opal,cleo2,aleph} is used to test the conservation of the vector 
 current (CVC).

 Theoretical calculations of the $e^+e^-\to e^+e^-,\pi^+\pi^-,\mu^+\mu^-$
 reactions cross sections play important role in the $e^+e^-\to\pi^+\pi^-$ 
 process measurements. They are necessary for luminosity
 measurements ($e^+e^-\to e^+e^-$ events), for the $e^+e^-\to\mu^+\mu^-$
 background subtraction, for the radiative corrections and
 $e^+e^-\to\pi^+\pi^-$ detection efficiency determination. The 
 $e^+e^-\to\pi^+\pi^-,\mu^+\mu^-$ cross sections were calculated according to
 the formulae of Ref. \cite{arbuzqed,arbuzhad}, which take into account
 the photons radiation by the initial and final state particles and have
 accuracy of about 0.2\%.
    
 Recently it was found that the $e^+e^-\to\pi^+\pi^-$ and $\mu^+\mu^-$ Monte
 Carlo events generators used in the SND data analysis were not quite correct
 and understated the $e^+e^-\to\pi^+\pi^-$ and $e^+e^-\to\mu^+\mu^-$ cross
 sections by about 2.5\% and 1.5\% respectively.

 In this paper, in order to correct the error, the measured cross section
 $\sigma_0$ is multiplied by correction factors:
\begin{eqnarray}
\sigma=\sigma_0\cdot\delta_\pi\cdot\delta_\mu,
\end{eqnarray}
 where $\delta_\pi$ and $\delta_\mu$ are corrections due to mistakes in the
 $e^+e^-\to\pi^+\pi^-$ and $e^+e^-\to\mu^+\mu^-$ cross sections calculations
 respectively. The $\delta_\pi$ and $\delta_\mu$ coefficients were determined
 using the MCGPJ $e^+e^-\to\mu^+\mu^-$, $\pi^+\pi^-$ events generator 
 \cite{mcjpg}, which is based on the same approach \cite{arbuzqed,arbuzhad}.
 The applied corrections have not altered the systematic error value of the
 $e^+e^-\to\pi^+\pi^-$ cross section measurement, which is 1.3 \% for the
 energy region $\sqrt{s}\ge 420$ MeV and 3.2 \% for $\sqrt{s}<420$ MeV.
 
\begin{figure}
\begin{center}
\epsfig{figure= 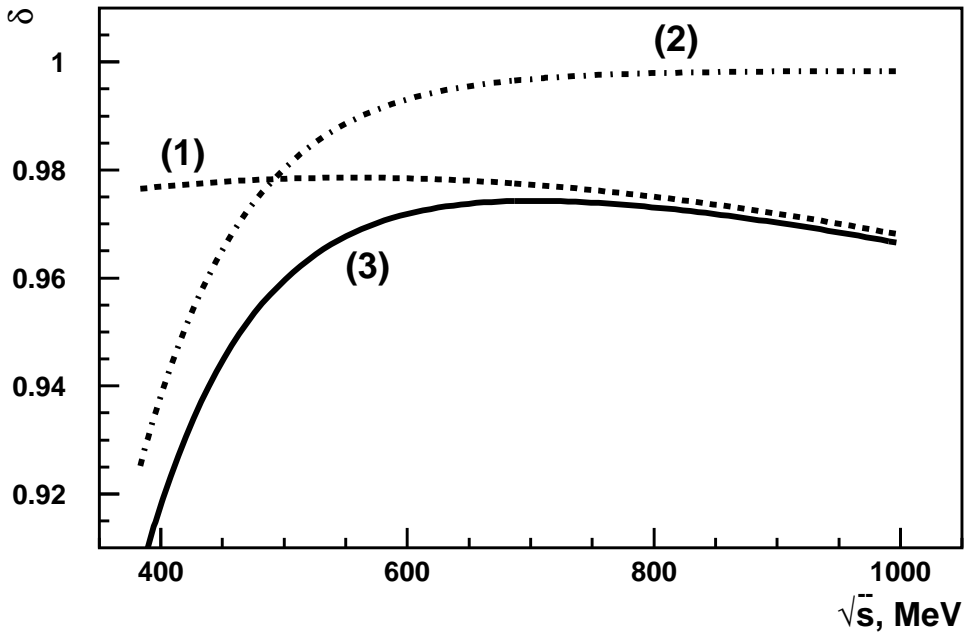,width=15.0cm}
\caption{Corrections $\delta=\delta_\pi\cdot\delta_\mu$ to the
         $e^+e^-\to\pi^+\pi^-$ cross section \cite{sndpipi}, which take into 
	 account the mistakes of the $e^+e^-\to\pi^+\pi^-$ (1) and
	 $e^+e^-\to\mu^+\mu^-$ (2) cross sections calculations and the total correction $\delta$ (3).}
\vspace{-0.5cm}
\label{nonpabku}
\end{center}
\end{figure}

 The energy dependence of the $\delta_\pi$ and $\delta_\mu$ is
 shown in Fig.\ref{nonpabku}. The error in the $e^+e^-\to\mu^+\mu^-$ cross
 section calculation is significant only in the energy region  $\sqrt{s}<500$
 MeV. The corrected values of the $e^+e^-\to\pi^+\pi^-$ cross section
 $\sigma_{\pi\pi}(s)$, of the form factor
\begin{eqnarray}
|F_\pi(s)|^2={{3s}\over{\pi\alpha^2\beta^3}}\sigma_{\pi\pi}(s),
       \mbox{~~} \beta=\sqrt{1-4m_{\pi}^2/s}
\end{eqnarray}
 and of the bare cross section (the cross section without vacuum polarization
 contribution but with the final state radiation taken into account) are
 listed in Table \ref{tab1}. The cross section decreased by two systematic
 errors in average. The results presented in Table \ref{tab1} supersede the
 results quoted in the original work \cite{sndpipi} Table 1.
 
\begin{figure}
\begin{center}
\epsfig{figure=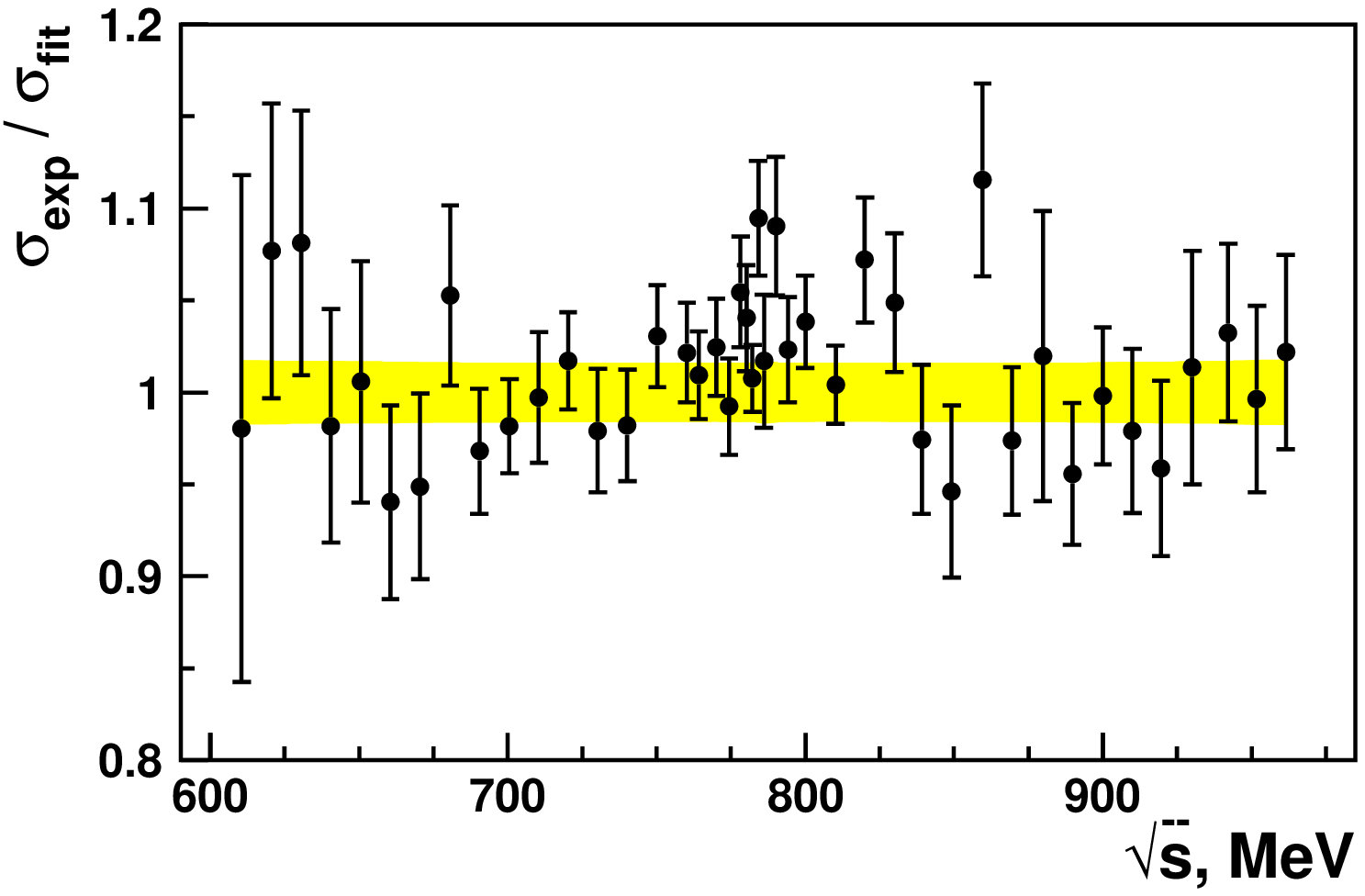,width=15.0cm}
\caption{The ratio $\sigma_{exp}/\sigma_{fit}$ of the $e^+e^-\to\pi^+\pi^-$
         cross section measured by CMD-2 \cite{kmd2} to the SND fit curve. The 
	 shaded area shows the joint systematic error.}
\label{cpacmd}
\epsfig{figure=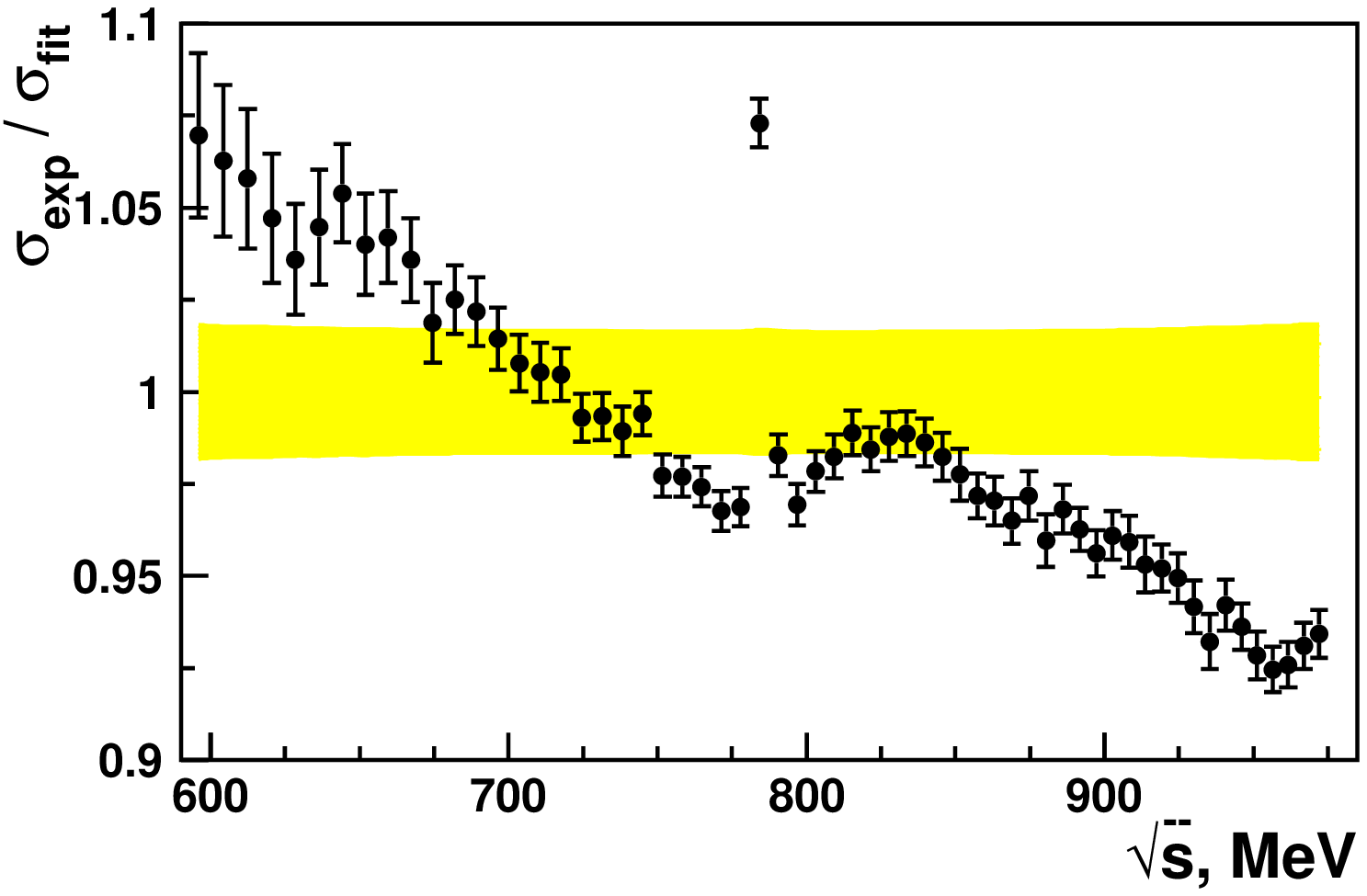,width=15.0cm}
\caption{The ratio $\sigma_{exp}/\sigma_{fit}$ of the $e^+e^-\to\pi^+\pi^-$
         cross section measured by KLOE \cite{kloe} to the SND fit curve.
	 The shaded area shows the joint systematic error.}
\label{cpakloe}
\end{center}
\end{figure}

 The comparison of the obtained cross section with CMD-2 \cite{kmd2} and KLOE
 \cite{kloe} measurements is shown in Figs.\ref{cpacmd},\ref{cpakloe}. The 
 CMD-2 result exceeds the SND data by $1.4\pm 1.5\%$ in average. Here the error
 includes both systematic and statistical uncertainties. The uncorrected SND
 cross section exceeds the CMD-2 one by the same value.
 In the KLOE experiment at the DA$\Phi$NE $\phi$-factory the form factor 
 $|F_\pi(s)|^2$ was measured using  ``radiative return'' method with the
 systematic error of 0.9 \% \cite{kloe}.  In Ref.\cite{kloe} the bare form 
 factor values are listed. So in order to compare the KLOE result with the SND
 one,  the form factor was appropriately dressed  by us. The results of this 
 comparison are shown in Fig.\ref{cpakloe}. The difference between SND and
 KLOE data is energy dependent. The point that jumped out is situated in the
 region of the sharp rise of the cross section due to the $\rho-\omega$
 interference. The KLOE measurement is in conflict with the SND result as well
 as with the CMD-2 one.
 
 The cross section was fitted as described in the original work \cite{sndpipi}.
 The fit results together with their deviation from the previous outcomes
 \cite{sndpipi} (in units of measurement errors) are listed in Table 
 \ref{tab2}. These values supersede the results of the previous work 
 \cite{sndpipi}. All parameters except $\sigma(\rho\to\pi^+\pi^-)$,
 $B(\rho\to e^+e^-)\times B(\rho\to\pi^+\pi^-)$ and $\Gamma(\rho\to e^+e^-)$
 changed by less than 0.5 error values, while $\sigma(\rho\to\pi^+\pi^-)$, 
 $B(\rho\to e^+e^-)\times B(\rho\to\pi^+\pi^-)$ and $\Gamma(\rho\to e^+e^-)$ --
 by less than two errors. The discussion of the parameters and conclusions 
 made in the original work \cite{sndpipi} are still valid.
 
\begin{figure}
\begin{center}
\epsfig{figure=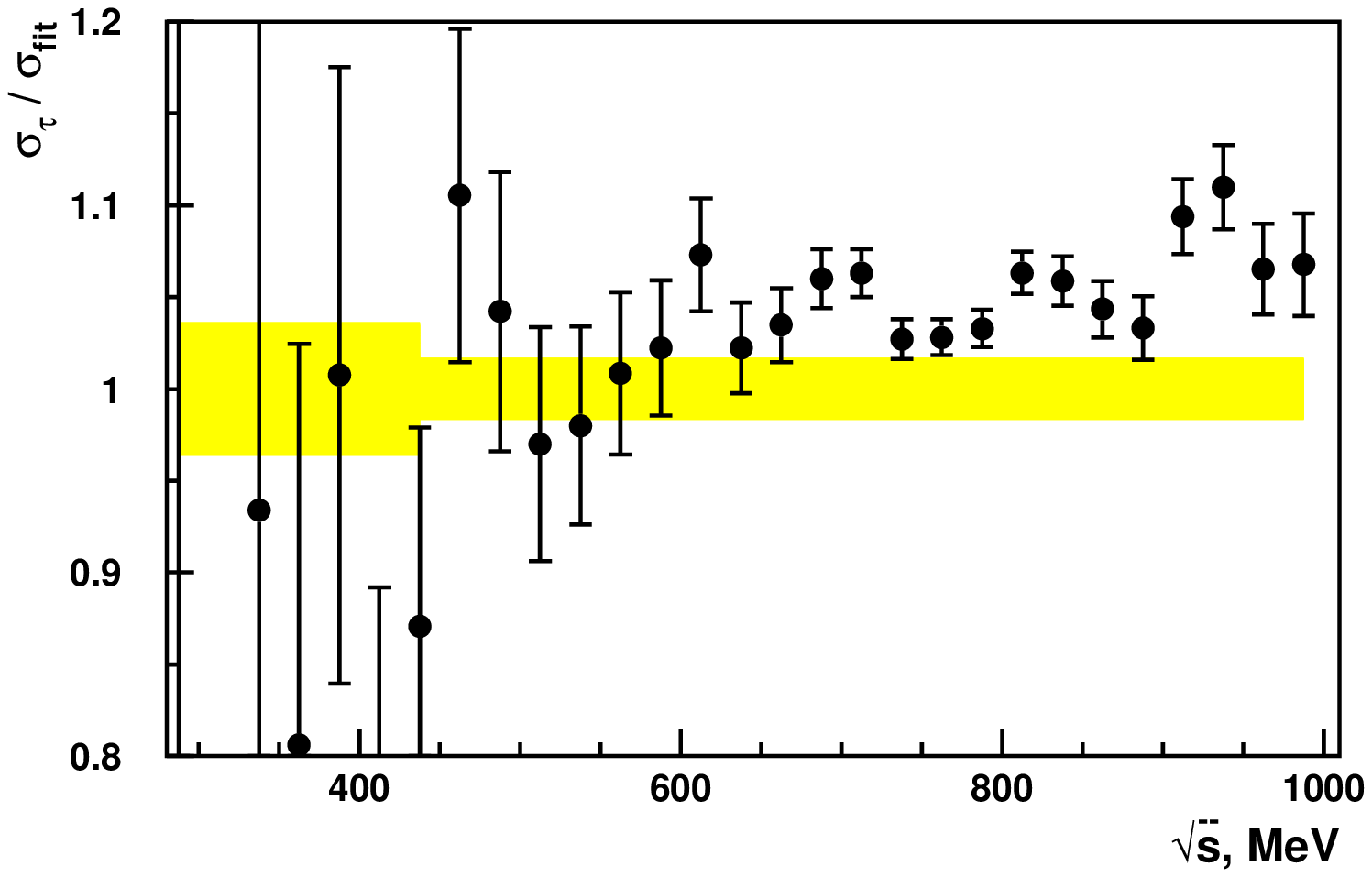,width=15.0cm}
\caption{The ratio of the $\sigma_\tau/\sigma_{fit}$ of the
         $e^+e^-\to\pi^+\pi^-$ cross section calculated from the
	 $\tau^-\to\pi^-\pi^0\nu_{\tau}$ decay spectral function measured by
	 CLEOII \cite{cleo2} to the isovector part of the $e^+e^-\to\pi^+\pi^-$
	 cross section corrected in this work. The shaded area shows the
         joint systematic error.}
\label{tau3}
\vspace{-1cm}
\epsfig{figure=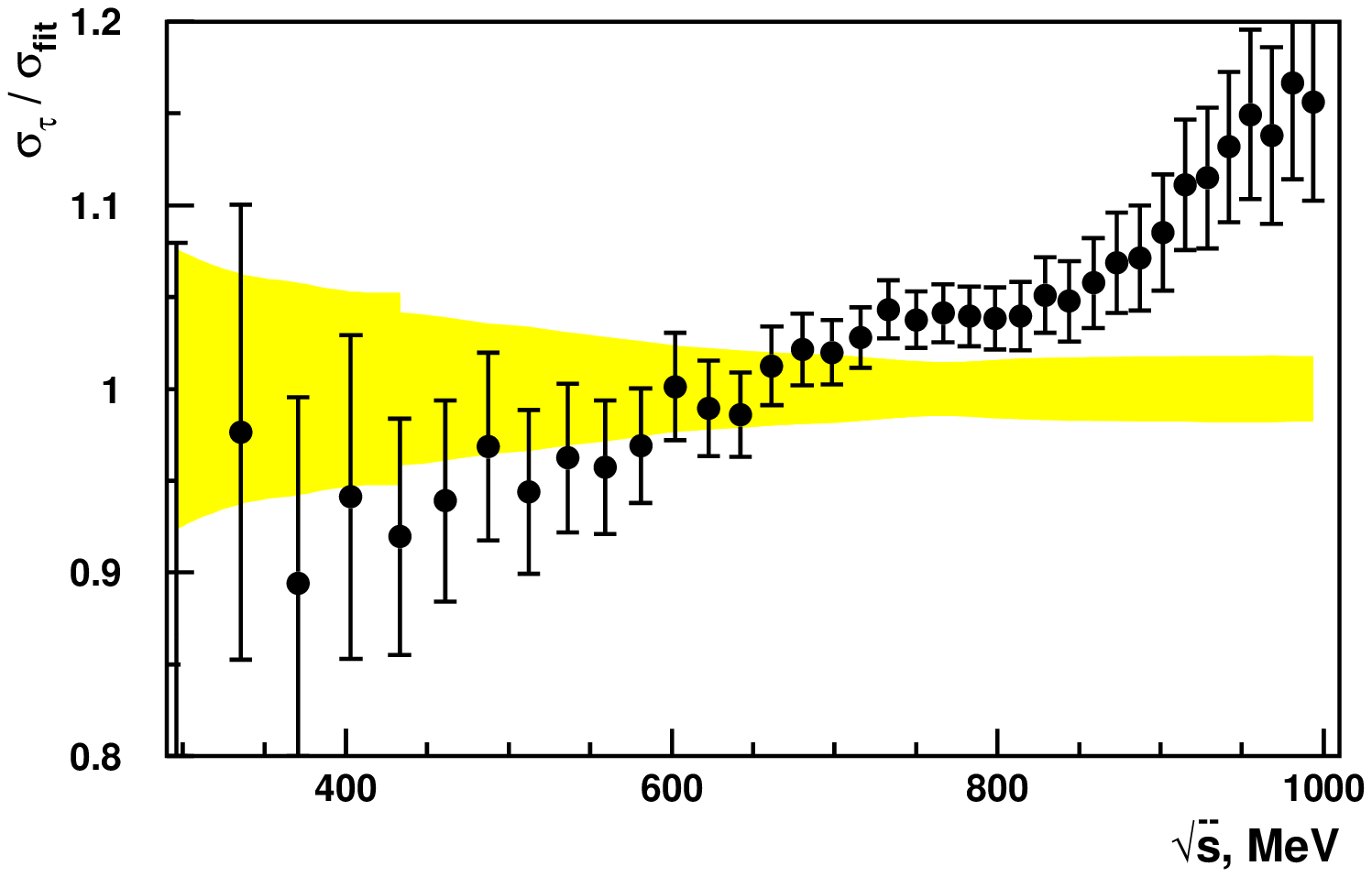,width=15.0cm}
\caption{The ratio of the $\sigma_\tau/\sigma_{fit}$ of the
         $e^+e^-\to\pi^+\pi^-$ cross section calculated from the
	 $\tau^-\to\pi^-\pi^0\nu_{\tau}$ decay spectral function measured by
	 ALEPH \cite{aleph} to the isovector part of the $e^+e^-\to\pi^+\pi^-$
	 cross section corrected in this work. The shaded area shows the
         joint systematic error.}
\label{tau4}
\end{center}
\end{figure}

\begin{figure}
\begin{center}
\epsfig{figure=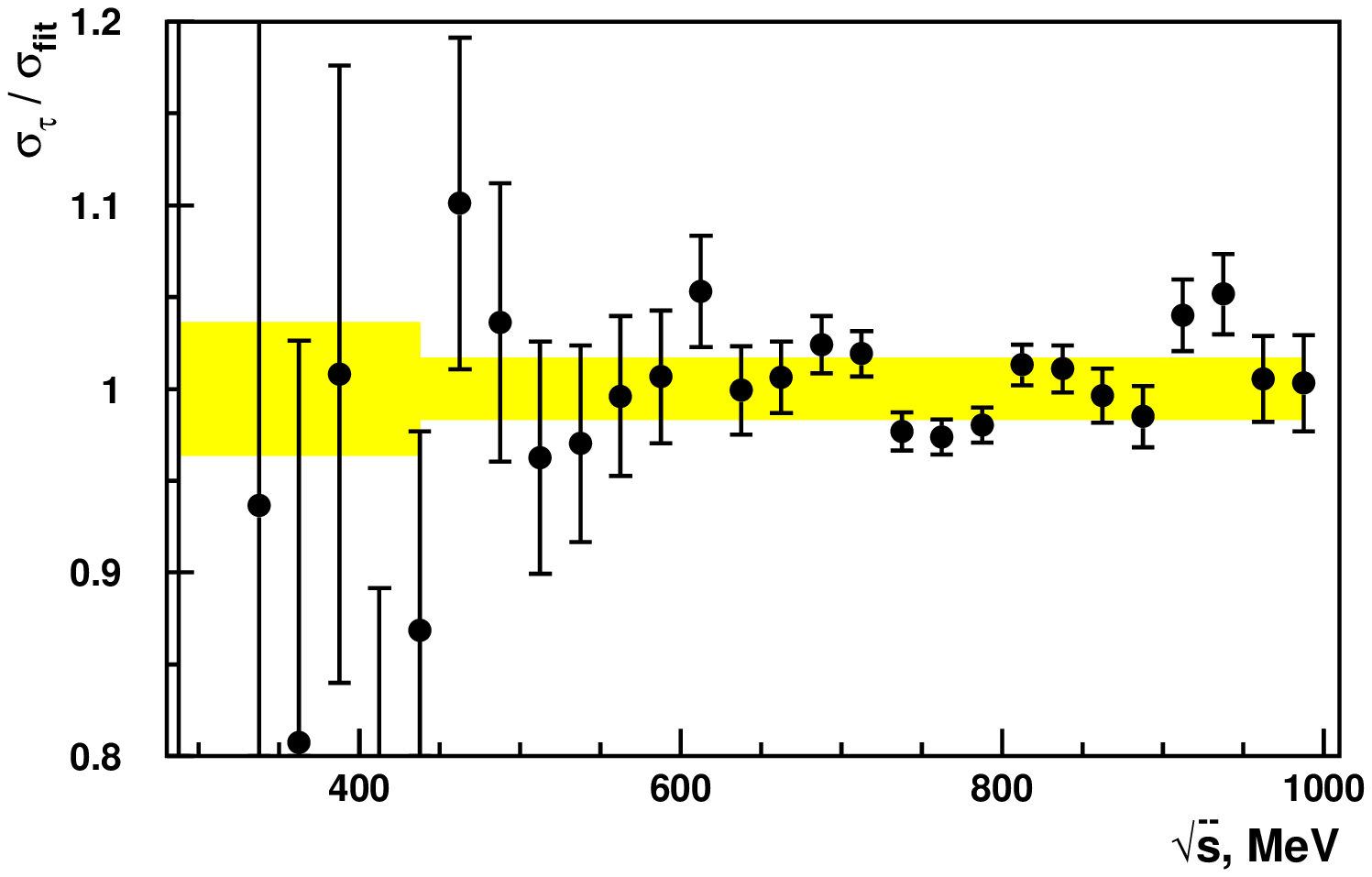,width=15.0cm}
\vspace{-0.5cm}
\caption{The ratio of the $\sigma_\tau/\sigma_{fit}$ of the
         $e^+e^-\to\pi^+\pi^-$ cross section calculated from the
	 $\tau^-\to\pi^-\pi^0\nu_{\tau}$ decay spectral function measured by
	 CLEOII \cite{cleo2} to the isovector part of the $e^+e^-\to\pi^+\pi^-$
	 cross section corrected in this work when the vacuum polarization
         contribution is not extracted from the SND data. The shaded 
	 area shows the joint systematic error.}
\label{tau1}
\vspace{-1cm}
\epsfig{figure=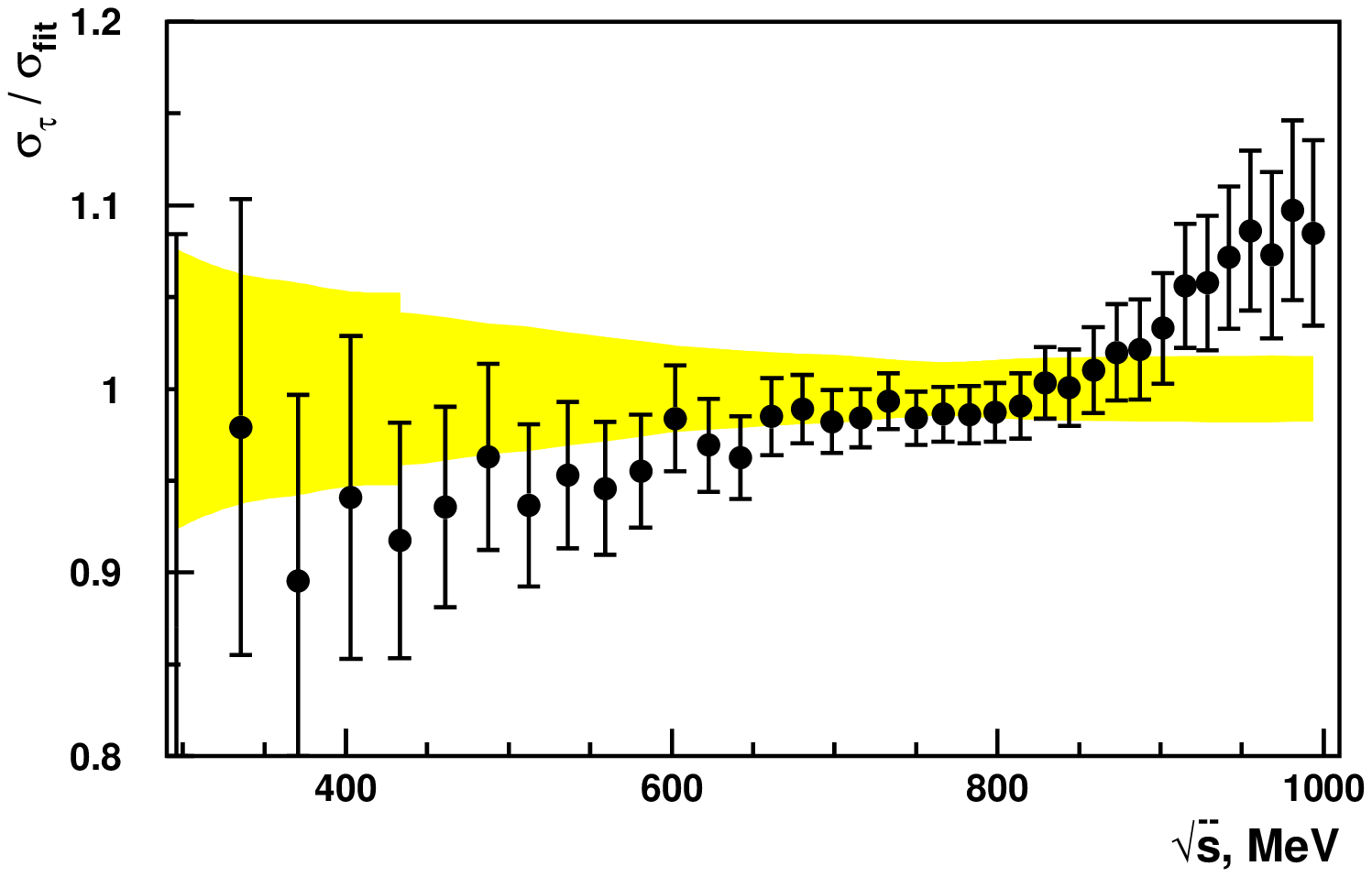,width=15.0cm}
\vspace{-0.5cm}
\caption{The ratio of the $\sigma_\tau/\sigma_{fit}$ of the
         $e^+e^-\to\pi^+\pi^-$ cross section calculated from the
	 $\tau^-\to\pi^-\pi^0\nu_{\tau}$ decays spectral function measured by
	 ALEPH \cite{aleph} to the isovector part of the $e^+e^-\to\pi^+\pi^-$
	 cross section corrected in this work. when the vacuum polarization
	 contribution is not extracted from the SND data. The shaded 
	 area shows the joint systematic error.}
\label{tau2}
\end{center}
\end{figure}

 The comparison of the $e^+e^-\to\pi^+\pi^-$ cross section obtained under the
 CVC hypothesis from the $\tau$-lepton spectral function from the
 $\tau^-\to\pi^-\pi^0\nu_{\tau}$ decay \cite{aleph,cleo2} with the isovector
 part of the cross section measured by SND is shown in Fig.\ref{tau3},
 \ref{tau4}. In order to compare with the $\tau$ spectral function, the
 radiative correction $S_{EW}=1.0198\pm 0.0006$ \cite{aleph,cleo2,radtau} was
 applied. The $e^+e^-\to\pi^+\pi^-$ cross section was undressed from the
 vacuum polarization, the contribution from the $\omega\to\pi^+\pi^-$ decay was
 excluded and correction for the $\pi^\pm$ and $\pi^0$ mass difference was 
 applied. As a result one can see the picture
 well known from Ref.\cite{eetau,aleph}. It is interesting that the difference
 between $e^+e^-$ and $\tau$ data is approximately equal to the value of the
 accepted vacuum polarization contribution to
 the $e^+e^-$ annihilation. The comparison of $\tau$ data with dressed 
 $e^+e^-\to\pi^+\pi^-$ cross section is shown in Fig.\ref{tau1}, \ref{tau2}.

 The authors are grateful to G.V. Fedotovich, F.V. Ignatov, G.N. Shestakov,
 A.L. Sibidanov for useful discussions. The work is supported in part by 
 grants Sci.School-905.2006.2, RFBR 04-02-16181-a, 04-02-16184-a, 
 05-02-16250-a, 06-02-16192-a.

\begin{table}
\caption{The results of the $e^+e^-\to\pi^+\pi^-$ cross section measurements.
         $\sigma_{\pi\pi}$ and
	 $|F_\pi|^2$ are the cross section and the  form factor of the 
	 $e^+e^-\to\pi^+\pi^-$ process, $\sigma^{pol}_{\pi\pi}$ is the
	 $e^+e^-\to\pi^+\pi^-$ undressed cross section without vacuum 
         polarization but with the final state radiation. 
         Only uncorrelated errors are shown.
	 The correlated systematic error $\sigma_{sys}$ is 1.3 \% for 
	 $\sqrt{s}\ge 420$ MeV and 3.2 \% for $\sqrt{s}< 420$ MeV.}
\label{tab1} 
\begin{tabular}[t]{cccccc}
$\sqrt[]{s}$ (MeV)&$\sigma_{\pi\pi}$(nb) &$|F_\pi|^2$&
$\sigma^{pol}_{\pi\pi}$(nb) \\ \hline
970.0&  76.68$\pm$ 1.79& 3.78$\pm$0.09&  75.06$\pm$ 1.75 \\
958.0&  91.33$\pm$ 1.96& 4.41$\pm$0.09&  89.22$\pm$ 1.91 \\
950.0& 101.52$\pm$ 1.93& 4.83$\pm$0.09&  99.07$\pm$ 1.88 \\
940.0& 115.14$\pm$ 1.57& 5.38$\pm$0.07& 112.25$\pm$ 1.53 \\
920.0& 147.78$\pm$ 5.15& 6.66$\pm$0.23& 143.57$\pm$ 5.00 \\
880.0& 246.38$\pm$ 2.80&10.30$\pm$0.12& 237.93$\pm$ 2.70 \\
840.0& 450.70$\pm$ 4.19&17.46$\pm$0.16& 433.39$\pm$ 4.03 \\
820.0& 622.54$\pm$ 5.54&23.19$\pm$0.21& 597.24$\pm$ 5.31 \\
810.0& 715.94$\pm$ 6.21&26.15$\pm$0.23& 685.26$\pm$ 5.94 \\
800.0& 822.66$\pm$ 7.05&29.46$\pm$0.25& 785.42$\pm$ 6.73 \\
794.0& 859.35$\pm$ 7.19&30.41$\pm$0.25& 815.84$\pm$ 6.83 \\
790.0& 855.55$\pm$16.98&30.04$\pm$0.60& 806.96$\pm$16.02 \\
786.0& 874.23$\pm$ 7.42&30.45$\pm$0.26& 820.44$\pm$ 6.96 \\
785.0& 887.68$\pm$ 8.81&30.86$\pm$0.31& 835.25$\pm$ 8.29 \\
784.0& 940.42$\pm$19.12&32.62$\pm$0.66& 890.94$\pm$18.11 \\
783.0&1022.45$\pm$10.99&35.40$\pm$0.38& 979.52$\pm$10.53 \\
782.0&1106.69$\pm$26.44&38.24$\pm$0.91&1073.85$\pm$25.66 \\
781.0&1161.62$\pm$10.84&40.06$\pm$0.37&1138.88$\pm$10.63 \\
780.0&1233.58$\pm$10.17&42.45$\pm$0.35&1220.00$\pm$10.06 \\
778.0&1314.33$\pm$ 9.78&45.05$\pm$0.34&1309.00$\pm$ 9.74 \\
774.0&1331.59$\pm$ 9.87&45.28$\pm$0.34&1326.86$\pm$ 9.83 \\
770.0&1302.16$\pm$ 9.67&43.92$\pm$0.33&1296.23$\pm$ 9.63 \\
764.0&1304.40$\pm$ 9.80&43.47$\pm$0.33&1297.32$\pm$ 9.75 \\
760.0&1308.40$\pm$10.08&43.26$\pm$0.33&1301.28$\pm$10.03 \\
750.0&1291.96$\pm$22.80&41.86$\pm$0.74&1288.31$\pm$22.74 \\
720.0&1060.14$\pm$ 7.11&32.31$\pm$0.22&1064.32$\pm$ 7.14 \\
\end{tabular}
\end{table}

\begin{table}
 Table I: (Continued) \\
\begin{tabular}[t]{cccccc}
$\sqrt[]{s}$ (MeV)&$\sigma_{\pi\pi}$(nb) &$|F_\pi|^2$&
$\sigma^{pol}_{\pi\pi}$(nb) \\ \hline
690.0& 764.53$\pm$ 8.31&21.92$\pm$0.24& 769.56$\pm$ 8.36 \\
660.0& 543.75$\pm$ 6.24&14.66$\pm$0.17& 546.05$\pm$ 6.27 \\
630.0& 398.61$\pm$ 8.73&10.11$\pm$0.22& 399.49$\pm$ 8.75 \\
600.0& 296.06$\pm$10.92& 7.08$\pm$0.26& 296.17$\pm$10.92 \\
580.0& 261.49$\pm$14.78& 6.01$\pm$0.34& 261.11$\pm$14.76 \\
560.0& 230.91$\pm$12.69& 5.12$\pm$0.28& 230.54$\pm$12.67 \\
550.0& 221.00$\pm$17.83& 4.81$\pm$0.39& 220.33$\pm$17.78 \\
540.0& 215.61$\pm$13.79& 4.62$\pm$0.30& 214.99$\pm$13.75 \\
530.0& 202.32$\pm$23.04& 4.26$\pm$0.49& 201.77$\pm$22.98 \\
520.0& 179.55$\pm$10.42& 3.72$\pm$0.22& 179.10$\pm$10.39 \\
510.0& 175.37$\pm$16.81& 3.58$\pm$0.34& 174.62$\pm$16.74 \\
500.0& 176.32$\pm$10.93& 3.55$\pm$0.22& 175.60$\pm$10.89 \\
480.0& 165.60$\pm$ 9.72& 3.26$\pm$0.19& 165.01$\pm$ 9.69 \\
470.0& 143.61$\pm$13.28& 2.81$\pm$0.26& 143.14$\pm$13.24 \\
450.0& 140.47$\pm$14.24& 2.71$\pm$0.28& 139.82$\pm$14.17 \\
440.0& 114.75$\pm$15.51& 2.22$\pm$0.30& 114.26$\pm$15.44 \\
430.0& 109.25$\pm$12.54& 2.11$\pm$0.24& 108.83$\pm$12.49 \\
410.0& 125.06$\pm$18.92& 2.46$\pm$0.37& 124.70$\pm$18.87 \\
390.0& 116.37$\pm$21.78& 2.39$\pm$0.45& 116.17$\pm$21.74 \\
\hline
\end{tabular}
\end{table}

\begin{table}
\begin{center}
\caption{The $\rho$ and $\omega$ mesons parameters measured in this work. 
         In the third column the parameters deviations from the original work 
	 results \cite{sndpipi} in units of measurement errors are listed.
	 \bigskip}
\label{tab2}
\begin{tabular}[t]{llc}
 Parameters & Values & Deviations \\
\hline
 $m_\rho,$ MeV&774.6$\pm$0.4$\pm$0.5&0.5 \\
 $\Gamma_\rho,$ MeV&146.1$\pm$0.8$\pm$1.5&0.2 \\
 $\sigma(\rho\to\pi^+\pi^-),$ nb&1193$\pm$7$\pm$16&1.6 \\
 $B(\rho\to e^+e^-)\times B(\rho\to\pi^+\pi^-)$&
 $(4.876\pm 0.023\pm0.064)\times 10^{-5}$&1.6\\
 $\Gamma(\rho\to e^+e^-)$, keV&$7.12\pm 0.02\pm0.11$&1.7\\
 $\sigma(\omega\to\pi^+\pi^-)$, nb&29.3$\pm$1.4$\pm$1.0&0.3\\
 $B(\omega\to e^+e^-)\times B(\omega\to\pi^+\pi^-)$&
 $(1.225\pm 0.058\pm0.041)\times 10^{-6}$&0.3\\
 $\phi_{\rho\omega}$, degrees&113.7$\pm$1.3$\pm$2.0&0.1 \\
\hline
\end{tabular}
\end{center}
\end{table}


\end{document}